\documentclass{ceurart}

\usepackage{amsmath}
\usepackage{amssymb}
\usepackage{microtype}
\usepackage{subcaption}
\usepackage{soul}
\definecolor{hlpurple}{RGB}{221,190,255}

\begin{document}

\copyrightyear{2026}
\copyrightclause{Copyright for this paper by its authors.
 Use permitted under Creative Commons License Attribution 4.0
 International (CC BY 4.0).}

\conference{}

\title{Integrity of peer-to-peer distributed LLM inference under malicious nodes}

\author[1]{Mert Cihangiroglu}[%
 email=mert.cihangiroglu01@universitadipavia.it,
]
\author[1]{Antonino Nocera}[%
 email=antonino.nocera@unipv.it,
]
\address[1]{DCALab, Department of Electrical, Computer and Biomedical Engineering,
 University of Pavia, Italy}

\begin{abstract}
Peer-to-peer distributed inference executes a Large Language Model (LLM) on pooled consumer hardware by spreading its layers across many nodes. Every request passes through nodes that are owned and controlled by multiple independent parties. However, in this setting, any party can tamper with the output of its layers to corrupt the end result. Recomputing the forward pass on trusted hardware can catch this, but it introduces additional computational cost. The scientific literature includes several prior integrity-checking approaches, such as known-answer traps for image classifiers and cryptographic commitments. However, these solutions test only the exact correctness and do not account for the ordinary variation that may arise between benign nodes. In this paper, we propose a method that checks the output integrity by measuring the variation in the activations that each node passes to the next. A peer who wants to use the network selects a small set of secret canary inputs whose correct activations are known in advance and mixes them into regular traffic. Because the peers cannot tell a canary from a real query, any tampering node corrupts them as well. The deviation from the known reference then reveals malicious activity: benign nodes exhibit only minor variation from hardware-induced noise, whereas tampered nodes deviate far more. We treat the identification of malicious nodes as a probabilistic test that separates two drift distributions, without relying on a fixed threshold. We study 408 configurations with metrics and success criteria fixed before any experiment ran; the detector reaches AUROC 1.0, correctly ranking the malicious shard above every benign shard on every canary in every configuration.
\end{abstract}


\maketitle

\section{Introduction}

A peer-to-peer distributed LLM inference pipeline assigns successive layers of a
model to different untrusted volunteer devices. A single forward pass therefore
hops across many machines a single peer does not control, and any one of those
peers can return tampered activations. Because each peer sees only an
intermediate hidden state, the prompter does not have a scalable way to determine which hop
corrupted the final answer, or it is corrupted at all. Detecting a malicious peer among benign ones is the integrity problem we study in this paper.

The problem we study is timely and crucial since the peer-to-peer pipeline is a strategy for running large models on pooled consumer hardware. Petals~\cite{petals2023} demonstrated the deployment. The authors built fault-tolerant inference algorithms and load-balancing protocols that automatically assign devices to maximize total throughput. Their work handles crash faults but not adversaries that stay online and maliciously alter activations. The security literature for this threat has previously addressed prompt privacy: the prompt inference attack of~\cite{promptinference2025} warns that distributing LLM layers across participants requires transmitting intermediate
outputs, which can expose the original input prompts. Output integrity under a malicious peer is the unexplored half of that threat surface.

Detection of the malicious peer is difficult because benign shards already deviate from any precomputed reference for various reasons. Heterogeneous hardware, float rounding, and algebraic reordering
can cause a benign fp16 activation to differ from an fp32 reference even when there is no malicious peer.Low-precision execution and floating-point non-associativity are the measured sources of this benign drift ~\cite{dettmers2022gpt3,shanmugavelu2024impacts}. If the smallest tampering that still degrades accuracy produces a mismatch indistinguishable from this benign float noise, no threshold separates
malicious from benign peers, and the detector collapses to random coin flip. Fully recomputing the pass at higher cost solves this problem. We aim to replace the expensive detection with a cheap one.

The closest prior trap-based defense is Fusion~\cite{fusion2022}, and the
closest activation-level verification mechanism is TOPLOC~\cite{toploc2025};
Section~\ref{sec:related} places both against the broader literature. Our work
transplants the known-answer canary trap onto the intermediate activations of a
multi-hop LLM forward pass. We frame detection as a data-driven statistical
test on activation data rather than a cryptographic commitment or a redundant-compute guaranty.

We simulate a multi-stage LLM pipeline with precomputed fp32 canary references,
benign fp16 noise, and persistently malicious peers. We run a
canary-versus-reference mismatch detector and score it by AUROC against a random
baseline and a redundant-recompute ceiling. Across a pre-registered factorial of
216 configurations our detector reaches AUROC 1.000 wherever the benign
noise floor stays below the tamper magnitude, a gain of 0.500 over the
0.50 random baseline that clears the pre-registered 0.40 threshold, with the
scrambled-label control collapsed to chance. Two substudies deliberately raise
that floor to meet the tamper and locate the point where detection degrades.

\paragraph*{Summary of Contributions.}
\begin{itemize}
\item We formulate malicious-shard detection in a peer-to-peer LLM inference pipeline as a threshold-free statistical test on intermediate activations, implemented by transplanting the known-answer canary trap from two-party secure computation onto a multi-hop forward pass.
\item We pin a measured random-detector lower bound near 0.50 and a
 redundant-recompute upper bound of 1.0 on the same canary inputs, framing
 detection against the cost tier it aims to replace.
\item We report a canary detector that reaches AUROC 1.000 across the
 factorial on both architectures,
 both datasets, and all six tampering strategies, a gain of 0.500 over the
 random baseline, with a collapsed scrambled-label control, and we locate the
 benign-noise floor at which detection breaks down.
\end{itemize}

\section{Related Work}
\label{sec:related}

\paragraph{Known-answer traps.}
The closest defense to ours is Fusion~\cite{fusion2022}. A client mixes public
samples whose correct outputs it already knows into its own queries, so a server
that departs from the protocol returns something the client can catch. Fusion
assumes a two-party secure computation over an image classifier: one server, one
decision, and one output to check. A peer-to-peer LLM pipeline breaks all three
assumptions. The forward pass crosses many machines, the quantity being checked is
an intermediate hidden state rather than a label, and detection has to name which
hop cheated. We keep the trap and rebuild the test around those differences,
treating the canary construction as public in the sense of
Kerckhoffs~\cite{kerckhoffs1883} so that security rests on which queries are
canaries and not on the design being secret.

\paragraph{Verifiable inference.}
A larger body of work proves that an untrusted server ran the model it claimed to
run. SafetyNets uses an interactive proof~\cite{ghodsi2017safetynets}, Slalom
leans on trusted hardware~\cite{tramer2018slalom}, and later systems use
zero-knowledge proofs~\cite{kang2022scaling}, optimistic fraud
proofs~\cite{conway2024opml}, or proofs about the training
compute~\cite{jia2021proof}. TOPLOC~\cite{toploc2025} is the closest of these to
our setting: it hashes intermediate activations with a locality-sensitive scheme
and catches unauthorized model or precision changes at 100\% accuracy. What unites
them is cost. Each one reruns the computation, re-derives it cryptographically, or
requires a trusted execution environment on every node. TOPLOC also commits to a
hash rather than measuring how far an activation drifted, so it returns a match or
a mismatch instead of a ranking. We read each activation once, never recompute,
and report a score rather than a verdict.

\paragraph{Peer-to-peer LLM serving.}
Petals~\cite{petals2023} showed that a 176B model can be served on pooled
consumer hardware, with fault-tolerant inference and load balancing that assign
devices to maximize throughput. Related systems extend the paradigm; in particular, SWARM
parallelism trains large models over unreliable, poorly connected
devices~\cite{ryabinin2301swarm}, FlexGen pushes high-throughput generation onto a
single GPU~\cite{sheng2023flexgen}, and EdgeShard~\cite{zhang2024edgeshard},
Helix~\cite{mei2025helix}, and PipeEdge~\cite{hu2021pipeline} partition inference
across heterogeneous edge devices. All of them treat node failure as a crash; in particular, a
peer goes offline, and the scheduler routes around it. None of the existing approaches handles a peer that
stays online and returns wrong answers. Security work in this setting has so far
addressed the other half of the threat surface, prompt privacy, where transmitting
intermediate outputs can leak the original input~\cite{promptinference2025}.

\paragraph{Byzantine-robust distributed learning.}
Identifying a malicious participant among benign ones is a long-standing problem
in federated learning. Robust aggregation rules, such as
Krum~\cite{blanchard2017machine} and coordinate-wise median or trimmed
mean~\cite{yin2018byzantine} discard updates that are far from the consensus,
though these rules have known blind spots~\cite{guerraoui2018hidden}. FLTrust
bootstraps from a small clean root dataset~\cite{cao2020fltrust}, and
blockchain-based schemes distribute the aggregation
itself~\cite{blockdfl2022,balance2024}; the broader threat landscape is surveyed
in~\cite{lyu2020threats}. Strong adaptive attacks still defeat many of
them~\cite{fang2020local}. Two things separate that setting from ours. These
defenses operate on model updates during training rather than on activation
streams during inference, and nearly all of them assume a benign majority,
because they identify an outlier by comparing it against the crowd. Our test
compares each shard against a stored reference instead, which is why it does not
degrade as the malicious minority grows.

\paragraph{Numerical nondeterminism.}
Integrity checking is hard here because the benign nodes disagree. Low-precision
execution~\cite{dettmers2022gpt3} and the non-associativity of floating-point
reduction across hardware~\cite{shanmugavelu2024impacts} are measured sources of
drift between two correct runs of the same model. Any exact-match check reads that
benign drift as tampering. Our detector is built around it. In particular, the benign floor is
the null distribution used as a reference by the test.

\section{Method}

In this section, we formally describe our method. At a high level, a verifier audits the untrusted peer-to-peer pipeline by planting known-answer canary queries that are indistinguishable from ordinary traffic. It precomputes each canary's exact fp32 stage activations offline, then compares them against the activations the live pipeline returns: any shard (peer) whose deviation exceeds the benign floating-point noise floor is flagged as tampering, and therefore malicious. Figure~\ref{fig:approach} illustrates this end-to-end check. We do so in five parts. Section~\ref{subsec:pipeline} defines the
multi-hop forward pass and the benign float-noise model. Section~\ref{subsec:threat} specifies
the adversary and the three tampering strategies that may be applied. Section~\ref{subsec:canary} sets out how the verifier stores fp32 references and routes canary
input through the live pipeline. Section~\ref{subsec:mismatch} defines the per-canary score
and casts detection as a threshold-free ranking test scored by AUROC. Section~\ref{subsec:bounds} fixes the random baseline, the scrambled-label control, and the redundant-recompute
ceiling that bracket the detector.
\subsection{Pipeline model}
\label{subsec:pipeline}
Let $k \in \mathbb{N}$ denote the pipeline depth. Stage $i \in [k] =
\{1,\ldots,k\}$ is assigned one transformer block $f_i: \mathbb{R}^{T \times
d} \to \mathbb{R}^{T \times d}$, where $T$ is the sequence length and $d$ is
the hidden dimension. Given an input query $x$, let $\mathbf{a}_0 \in
\mathbb{R}^{T \times d}$ be its token embedding. The \emph{clean fp32 reference
activations} are defined recursively:
\begin{equation}
 \mathbf{a}_i^* \;=\; f_i\!\left(\mathbf{a}_{i-1}^*\right), \quad i \in [k].
 \label{eq:clean}
\end{equation}
In the live pass, let $\mathbf{h}_i$ denote the live activation in stage $i$. Each benign stage $i \notin \mathcal{M}$, where $\mathcal{M}$ is the set of malicious stages defined in Section~\ref{subsec:threat}, produces
\begin{equation}
 \mathbf{h}_i^{\mathrm{ben}} \;=\; f_i\!\left(\mathbf{h}_{i-1}\right) +
 \boldsymbol{\delta}_i,
 \qquad
 \boldsymbol{\delta}_i \;\sim\; \mathcal{N}\!\left(\mathbf{0},\;
 \sigma_{hw}^2\,\mu_i^2\,\mathbf{I}\right),
 \qquad
 \mu_i \;=\; \frac{1}{Td}\bigl\|f_i(\mathbf{h}_{i-1})\bigr\|_1,
 \label{eq:benign}
\end{equation}
where $\sigma_{hw} > 0$ is the hardware noise level and $\mu_i$ is the mean
absolute activation of stage $i$. Equation~\eqref{eq:benign} models
heterogeneous-hardware nondeterminism. In particular, the per-element noise scales with the
activation magnitude through $\mu_i$, so the \emph{relative} displacement is
proportional to $\sigma_{hw}$ and independent of $d$ and $T$.

\subsection{Threat model}
\label{subsec:threat}
The adversary controls a set $\mathcal{M} \subseteq [k]$ of $c = |\mathcal{M}|$
colluding stages. At each malicious stage $i^* \in \mathcal{M}$, the adversary
replaces the benign computation with a tampering function $\tau_{i^*}:
\mathbb{R}^{T \times d} \to \mathbb{R}^{T \times d}$:
\begin{equation}
 \mathbf{h}_{i^*}^{\mathrm{mal}} \;=\; \tau_{i^*}\!\left(\mathbf{h}_{i^*-1}\right).
 \label{eq:tamper}
\end{equation}
Let $\mathbf{v} = f_{i^*}(\mathbf{h}_{i^*-1})$ denote the benign output before
tampering and $\epsilon > 0$ be the magnitude of tampering. We study three tampering
strategies, evaluated as six attacks by sweeping the Gaussian magnitude over four values:
\begin{align*}
 \text{Gaussian:}\ & \tau(\mathbf{v}) = \mathbf{v} + \epsilon \|\mathbf{v}\|_F \mathbf{U},\
 \mathbf{U} \sim \mathcal{N}(\mathbf{0}, \mathbf{I})/\|\mathcal{N}(\mathbf{0},\mathbf{I})\|_F,\
 \epsilon \in \{0.02, 0.05, 0.10, 0.20\};\\
 \text{Sign flip:}\ & \tau(\mathbf{v}) = -\mathbf{v};\\
  \text{Rescale:}\ & \tau(\mathbf{v}) = (1+\alpha)\mathbf{v},\quad \alpha = 0.10.
\end{align*}
These three strategies demonstrate ways of corrupting a deployed model's computation, ranging from bit-flip and rowhammer attacks that flip weight bits in memory~\cite{rakin2019bit,yao2020deephammer} to bounded adversarial weight perturbations~\cite{wu2020adversarial}.
Following Kerckhoffs's principle~\cite{kerckhoffs1883}, we treat the canary construction as public. We assume the adversary observes $\mathbf{h}_{i^*-1}$, knows $f_{i^*}$, tampers on every
query including canaries, but cannot distinguish a canary from a real query at
serving time and so cannot selectively spare the canaries.

\subsection{Reference precomputation and the canary trap}
\label{subsec:canary}
Let $\mathcal{C} = \{x^{(t)}\}_{t=1}^{n}$ be $n$ canary queries drawn from the
same distribution as the real input. Known-answer canaries are a standard strategy for probing model behavior, used to measure unintended memorization~\cite{carlini2019secret} and to watermark generated text~\cite{kirchenbauer2023watermark}; in our strategy, they carry an integrity signal. Before deployment, the verifier (the trusted party running the integrity check) runs the
full pipeline in fp32 and stores the reference activations
$\mathbf{a}_i^{*(t)}$ for all $i \in [k]$, $t \in [n]$ from~\eqref{eq:clean}.
At serving time, each canary is routed through the live pipeline in fp16,
producing live activations $\{\mathbf{h}_i^{(t)}\}_{i \in [k]}$. Benign stages
contribute float noise per~\eqref{eq:benign}; malicious stages inject tampering
per~\eqref{eq:tamper}. The verifier compares each live output to its stored
reference without re-executing any block.

\subsection{Relative-L2 mismatch and detection AUROC}
\label{subsec:mismatch}
For canary $t$ and stage $i$, the \emph{relative-L2 mismatch} is as follows.

\begin{equation}
 m_i^{(t)} \;=\; \frac{\bigl\|\mathbf{h}_i^{(t)} -
 \mathbf{a}_i^{*(t)}\bigr\|_F}{\bigl\|\mathbf{a}_i^{*(t)}\bigr\|_F}
 \label{eq:mismatch}
\end{equation}

Under~\eqref{eq:benign}, a benign stage satisfies $m_i^{(t)} = O(\sigma_{hw})$.
A malicious stage with Gaussian tamper of magnitude $\epsilon \gg \sigma_{hw}$
satisfies $m_{i^*}^{(t)} \approx \epsilon$, far above the benign floor.
Because no fixed threshold separates benign from malicious mismatches in
general, detection is cast as a ranking problem. The per-canary detection AUROC
(area under the receiver operating characteristic curve) is the probability that the malicious shard outranks a randomly drawn benign
shard:
\begin{equation}
 A^{(t)} \;=\; \frac{1}{k - |\mathcal{M}|} \sum_{\substack{j=1 \\ j \notin
 \mathcal{M}}}^{k} \mathbf{1}\!\left[m_{i^*}^{(t)} > m_j^{(t)}\right],
 \label{eq:auroc_canary}
\end{equation}
where $\mathbf{1}[\cdot]$ is the indicator function. The aggregate detection AUROC over all canaries is
\begin{equation}
 \widehat{A} \;=\; \frac{1}{n}\sum_{t=1}^{n} A^{(t)}.
 \label{eq:auroc_agg}
\end{equation}

Equation~\eqref{eq:auroc_agg} is equivalent to the Mann-Whitney $U$ statistic
normalized to $[0,1]$. Here, $\widehat{A} = 1$ means $m_{i^*}^{(t)}$ exceeded every
benign shard's mismatch on every canary. In our strategy we do not set any threshold; the
verifier reports $\widehat{A}$ as a score.

\section{Experimental Setup}
\label{sec:setup}
Our evaluation aims at verifying whether canary-based integrity detection reliably identifies a
malicious node across the conditions an adversary controls, namely: the model
architecture, the dataset, the tampering strategy, the pipeline depth, the
attacker's position, the number of colluding nodes, and the benign hardware
noise floor $\sigma_{hw}$. We evaluate a factorial of 216 configurations
together with five focused substudies totaling 192 additional configurations,
for 408 configurations in total. 

\subsection{Models}
\begin{figure}[t]
\centering
\includegraphics[width=0.82\linewidth]{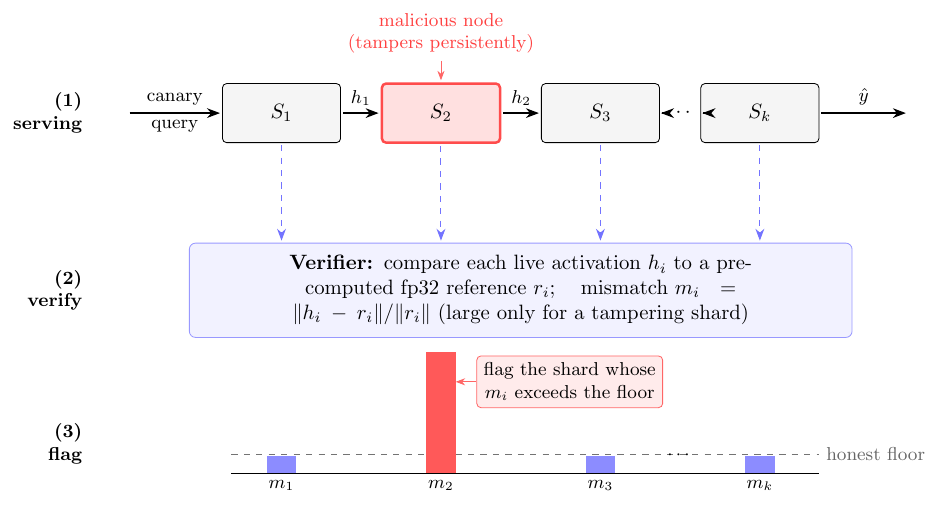}
\caption{The canary-trap integrity check. (1)~A canary query traverses $k$
untrusted shards of the peer-to-peer pipeline, one of which is malicious and
tampers persistently. (2)~The verifier compares each live stage activation
$\mathbf{h}_i$ to the precomputed fp32 reference $\mathbf{a}_i^*$ through the
relative-L2 mismatch $m_i$ from~\eqref{eq:mismatch}. (3)~The malicious shard's
mismatch $m_{i^*} \approx \epsilon$ exceeds the benign noise floor $m_j \approx
\sigma_{hw}$, flagging the tampering stage.}
\label{fig:approach}
\end{figure}

We instantiate each stage $f_i$ as one transformer block of either GPT-2
(124M parameters, 12 blocks, $d=768$) or Pythia-160M (GPT-NeoX architecture,
160M parameters, $d=768$), so a model with $k$ blocks executes as a $k$-hop
peer-to-peer forward pass.
The depth substudy adds a third model, Pythia-410M (GPT-NeoX architecture,
410M parameters, 24 blocks), to test whether the margin survives on a larger
model. It is used only there, in 18 of the 408 configurations; the factorial and
the other four substudies use GPT-2 and Pythia-160M. When a stage budget
$k$ is smaller than a model's block count, the audited pipeline is its first
$k$ blocks, so Pythia-410M runs the same $k \le 12$ grid as the other two.
Clean fp32 references are precomputed
per~\eqref{eq:clean}. The live pass runs in fp16 with $\sigma_{hw} = 0.002$
in~\eqref{eq:benign}. One or more stages are designated malicious and apply
$\tau_{i^*}$ from~\eqref{eq:tamper} to every query.
Figure~\ref{fig:approach} summarizes the pipeline.

\subsection{Datasets}
Canaries are drawn from the test split of two standard language-modeling
benchmarks: LAMBADA (the OpenAI variant), whose
last-word prediction task is designed so that the final token is recoverable only
from long-range context, and WikiText-2 (raw). Each configuration uses
$n = 100$ canaries, truncated to 48 tokens. Using two datasets that differ in
this way separates the detector's behavior from any single text distribution. In particular,
LAMBADA canaries are hard for the model, WikiText-2 canaries are ordinary prose.

\subsection{Attacks}
The three tampering strategies of~\eqref{eq:tamper} are evaluated as
six attacks. Gaussian perturbation is swept over four magnitudes,
$\epsilon \in \{0.02, 0.05, 0.10, 0.20\}$, because it is the only strategy
with a free magnitude parameter. This sweep is what locates the detector's
sensitivity floor, and it brackets the range from a tamper that costs almost no
accuracy to one that costs several points. Sign flip and the
$\alpha = 0.10$ rescale have no such parameter and contribute one attack
each. The three are chosen in such a way as to comprise all the axes a detector can fail on rather than to
enumerate realistic exploits. Specifically, sign flip changes direction and magnitude, the
rescale changes magnitude alone, and the Gaussian perturbation changes both by a
controlled amount. The
magnitudes and $\alpha$ are fixed in the pre-registration, before any
configuration is actually run.

\subsection{Experimental design}
The main factorial places the attacker on a single mid-pipeline shard ($i^* =
\lfloor k/2 \rfloor$, $c=1$) and crosses two model architectures (GPT-2;
Pythia-160M), two standard datasets (LAMBADA last-word prediction; WikiText-2),
six attacks ($\tau$ from~\eqref{eq:tamper} with $\epsilon \in
\{0.02, 0.05, 0.10, 0.20\}$, sign flip, and $\alpha=0.10$ rescale), three
detector settings (the full mechanism and two ablations), and three random seeds.
The product is $2 \times 2 \times 6 \times 3 \times 3 = 216$ configurations,
each evaluated on $n=100$ canaries.
A seed fixes the canary sample, the benign noise draw
of~\eqref{eq:benign}, and the random direction of the Gaussian tamper, so
the three seeds resample all of the stochastic parts of a configuration
together.
The first ablation, \emph{noisy reference},
removes the stored fp32 reference and compares the live output to a separately
drawn benign pass instead. The second ablation, \emph{cosine metric}, replaces
the relative-L2 mismatch in~\eqref{eq:mismatch} with $\tilde{m}_i^{(t)} = 1 -
{\mathbf{h}_i^{(t)} \cdot
\mathbf{a}_i^{*(t)}}/\bigl(\|\mathbf{h}_i^{(t)}\|_F\|\mathbf{a}_i^{*(t)}\|_F\bigr)$.

The cosine distance measures only the angle between two vectors, so it ignores any change in their length. The rescale strategy multiplies the activation by a positive constant, which changes its length but not its direction. The tampered stage therefore has cosine mismatch $\tilde{m}_{i^*}^{(t)} = 0$ and stops outranking the benign stages, so detection should break on rescale and not on other attack. We include the ablation to test this hypothesis. Specifically, the detection drop that appears only on the rescale, and only under the cosine metric, confirms that sensitivity to perturbation magnitude carries the relevant signal to catch the tampering.

Five substudies vary one factor around a shared anchor (the two models,
LAMBADA, Gaussian tamper at $\epsilon=0.05$, full mechanism, three seeds),
adding 192 configurations. The \emph{signal-to-noise} substudy increases
$\sigma_{hw}$ from 0.002 to 0.2 to locate the breaking point predicted
by~\eqref{eq:auroc_canary}. In practice, $\widehat{A}$ degrades toward 0.50 when $\sigma_{hw}
\approx \epsilon$. The \emph{depth} substudy varies $k \in \{2,4,6,8,10,12\}$
for all three models, including
Pythia-410M, whose first 12 transformer blocks form the audited pipeline. The
\emph{position} substudy varies $i^*$
across five fractions of the pipeline. The \emph{collusion} substudy increases
$c = |\mathcal{M}|$ from 1 to 4, with colluders spaced uniformly. The
\emph{frontier} substudy reduces $\epsilon$ from 0.10 to 0.001 at the fixed
floor and overlays detection with accuracy harm.

\subsection{Evaluation metrics}
\label{subsec:bounds}
 We report two quantities throughout the whole experimental evaluation. Detection is the per-canary AUROC
$\widehat{A}$ of~\eqref{eq:auroc_canary}, which asks whether the
malicious shard's mismatch outranks every benign shard's on the same canary, and
needs no threshold. Harm is the accuracy drop. Specifically, it represents the fall in last-token prediction
accuracy, in percentage points, between a clean pass and a tampered pass over the
same canaries, where a prediction counts as correct if the model's highest-scoring
next token is the true one. Both are reported against $\epsilon$ in Section~\ref{subsec:res-frontier}.

Three reference conditions bracket the detector. The \emph{random baseline}
replaces each $m_i^{(t)}$ with $\mathrm{Uniform}(0,1)$, giving
$\mathbb{E}[\widehat{A}] = 0.50$. The \emph{scrambled-label negative control}
retains the true mismatch values $\{m_i^{(t)}\}$ but replaces $i^*$ with a
randomly chosen stage $j^* \neq i^*$; if the separation arises from the canary
mechanism and not from artifacts of the measurement setup, $\widehat{A}$ should remain near 0.50. The \emph{redundant-recompute ceiling}
replaces each $m_i^{(t)}$ with the mismatch between live output and an
independent benign recompute of the same stage on the same input. This ceiling
matches the guarantee of full redundant recomputation at twice the compute, and
is the upper bound this detector aims to approach without paying that cost.

\subsection{Pre-registered protocol}
We fixed our evaluation protocol before running any experiment. Every criterion below was set in advance. Our primary metric is the gain $\widehat{A} - 0.50$, and we set the success threshold at 0.40, equivalently $\widehat{A} \geq 0.90$. The scrambled-label negative control (evaluated on its pooled mean) must satisfy
$\widehat{A}_{\mathrm{ctrl}} \in [0.45, 0.55]$, and the false-positive rate in
benign shards alone must remain at or below 5\%. The registered falsifier:
if the gain is at most 0.0, then $m_{i^*}^{(t)}$ is stochastically not greater
than $m_j^{(t)}$, benign nondeterminism overlaps the tamper signal, and the
canary trap is refuted along with its cheaper-than-recompute claim.

\section{Experimental Results}
\label{sec:results}

In this section, we report our experimental results. The complete detector separates the
malicious node from the benign shards, reaching
$\widehat{A} = 1.000$ at the modeled noise floor $\sigma_{hw} = 0.002$
on all 72 factorial configurations in which it is active, for
both architectures, both datasets and all six attacks. The separation is complete: on
every one of the 100 canaries per configuration, the malicious shard's mismatch is
larger than every benign shard's mismatch. The redundant-recompute ceiling also reaches
1.000, so canary detection matches that far more expensive defense while reading each
activation once instead of recomputing it. The random baseline stays at 0.50 (range
0.461 to 0.544), and the scrambled-label negative control collapses to 0.50 (range 0.448
to 0.549), so the signal comes from matching mismatches to the right shard, not from an
artifact of the measurement setup. These results come from 408 configurations, 216 in
the factorial and 192 in the substudies. The subsections that follow report the main
experiment, then the two ablation studies that identify which component carries the signal,
then the five substudies that map where the detector breaks down.
Table~\ref{tab:main} summarizes the factorial, and Figure~\ref{fig:detectors}
shows the per-canary AUROC against the three reference conditions.

\subsection{Detecting a single malicious node}
\label{subsec:res-harm}
In the main experiment, one shard in the middle of the pipeline is malicious and the
verifier has to pick it out from the benign ones. We repeat that setup for every
combination of the two models, the two datasets, the six attacks (the four Gaussian
perturbations at $\epsilon \in \{0.02, 0.05, 0.10, 0.20\}$, the sign flip and the
$\alpha = 0.10$ rescale), the three detector settings, and three seeds, which is the
216 configurations of Table~\ref{tab:main}.
In the 72 of them that use the full detector, the malicious shard's mismatch is larger
than every benign shard's mismatch on all 100 canaries, so $\widehat{A} = 1.000$
in every one (Figure~\ref{fig:detectors}).

Those six attacks differ sharply in how much damage they do to the answer, and
detection does not follow that damage (Table~\ref{tab:harm} and
Figure~\ref{fig:ablation}, right). First, the sign flip, which costs 26.8
accuracy points on average, more than any other strategy. Second, the Gaussian
perturbations, which scale with magnitude: 0.5 points at $\epsilon=0.02$, rising
to 9.6 points at $\epsilon=0.20$. Third, the $\alpha=0.10$ rescale, which
costs 0.5 points. $\widehat{A} = 1.000$ for all three. A rescale that barely
moves the answer is as visible to the detector as a sign flip that wrecks it.

\subsection{Detector ablations}
\label{subsec:res-ablation}

\begin{figure}[t]
\centering
\includegraphics[width=\linewidth]{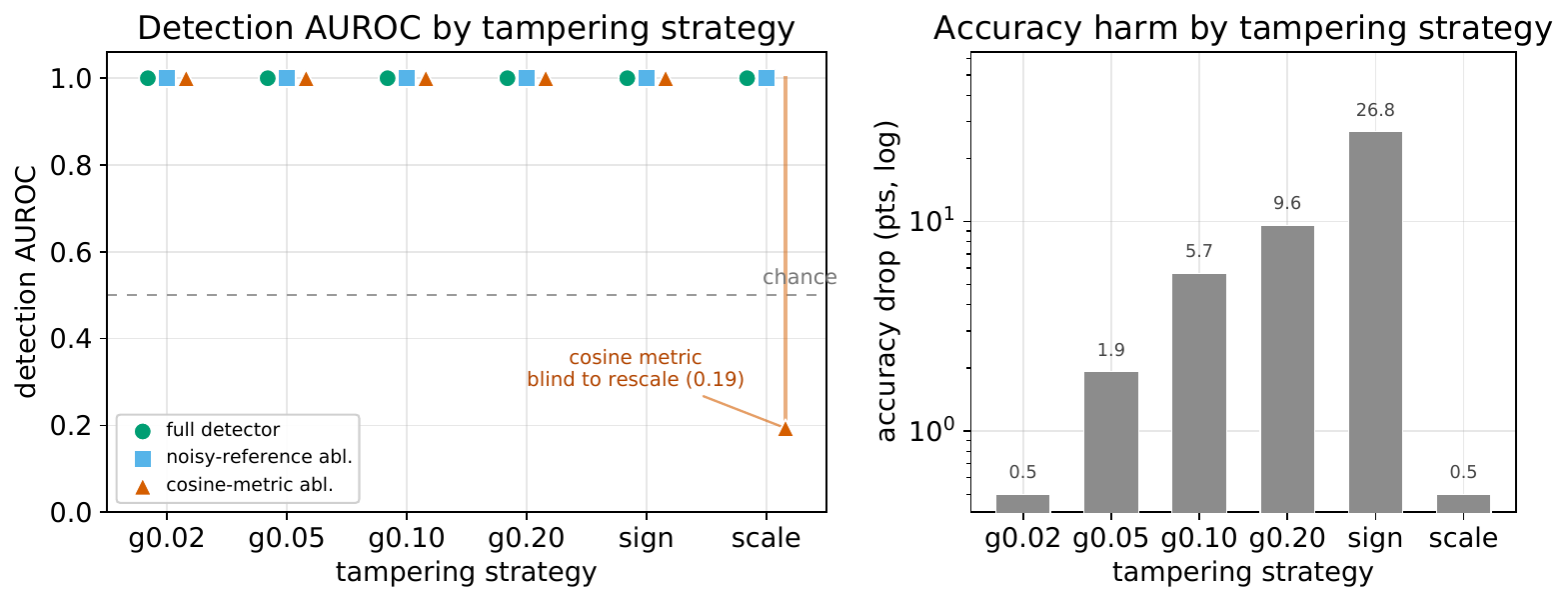}
\caption{Left: $\widehat{A}$ by tampering strategy for the full detector and the
two ablations; the cosine-metric ablation goes blind only on the pure rescale
($\widehat{A} = 0.19$), where scale invariance hides the magnitude-only change.
Right: harm of each strategy on a log axis. $\widehat{A} = 1.0$ regardless of
harm.}
\label{fig:ablation}
\end{figure}

Two ablations ask which part of the detector carries the signal
(Figure~\ref{fig:ablation}, left). First, we drop the stored fp32 reference and
compare the live pass against a separately drawn benign pass instead.
$\widehat{A}$ stays at 1.000. The stored reference is therefore not what
separates the two distributions; the float-noise gap is wide enough that even a
noisy reference finds it. Second, we swap relative-L2 for the cosine distance.
$\widehat{A}$ stays at 1.000 on five of the six attacks and falls to 0.194 on the
0.10 rescale, below the 0.50 chance line. That is the failure we mentioned when we
defined the ablation in Section~\ref{sec:setup}; in practice, the cosine distance reads direction
only, a pure rescale changes magnitude only, so the tampered stage stops standing out.
Relative-L2 in~\eqref{eq:mismatch} is the component that matters, and the rescale
is the one attack it only catches.

\subsection{Depth, position, and collusion substudies}
\label{subsec:res-robustness}

Next, we give the attacker more power. In particular, we allow deeper pipelines, any position in the
chain, and accomplices. None of these additional capabilities provide advantage to it (Figure~\ref{fig:robustness}). First, depth changes nothing: $\widehat{A} = 1.000$ for every $k$ from 2 to 12, on all three models including Pythia-410M. Benign noise grows with depth, so we expected some decay by $k = 12$, but none appears; $\epsilon/\sigma_{hw}$ stays wide enough at every depth we test. Second, the position does not make any difference. $\widehat{A} = 1.000$
wherever the malicious shard sits from the first stage through the last.Third, collusion does not help the attacker: $\widehat{A} = 1.000$ for $c = 1, 2, 3, 4$. Equation~\eqref{eq:auroc_canary} scores each colluder on its own against the benign shards, so nothing degrades as the malicious minority grows.

\subsection{Signal-to-noise substudy}
\label{subsec:res-snr}

\begin{figure}[t]
\centering
\begin{subfigure}[t]{0.49\linewidth}
  \centering
  \includegraphics[width=\linewidth]{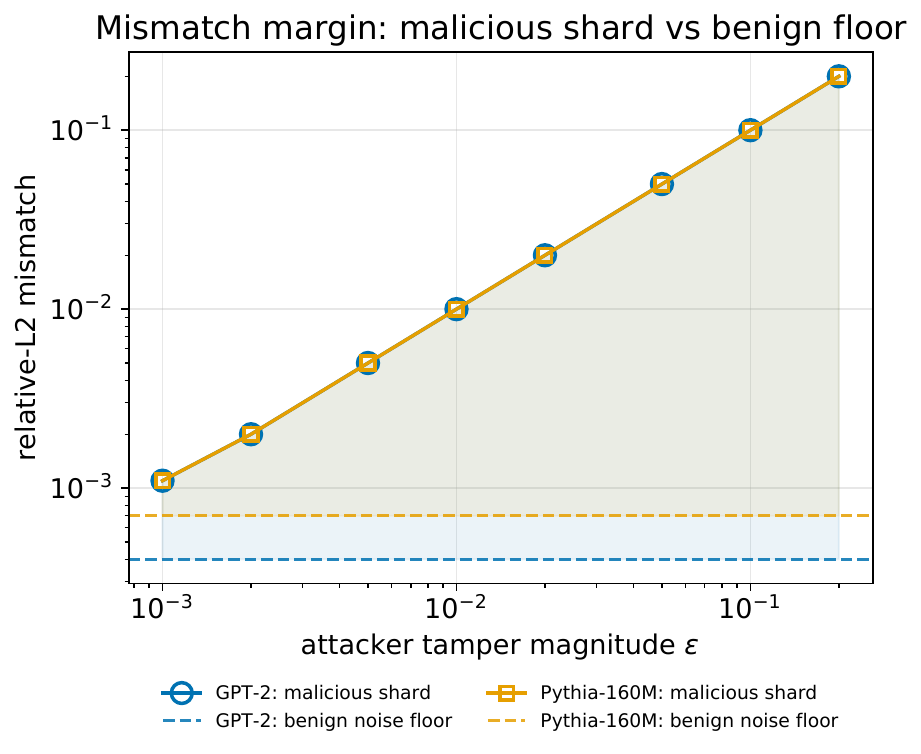}
  \caption{Why detection works. The malicious shard's mismatch $m_{i^*}^{(t)}$
  (markers) tracks $\epsilon$ along the diagonal; the benign floor
  $m_j^{(t)} \approx \sigma_{hw}$ (dashed) stays far below. Both axes are
  logarithmic; the shaded gap is the margin $\epsilon/\sigma_{hw}$.}
  \label{fig:mechanism}
\end{subfigure}\hfill
\begin{subfigure}[t]{0.49\linewidth}
  \centering
  \includegraphics[width=\linewidth]{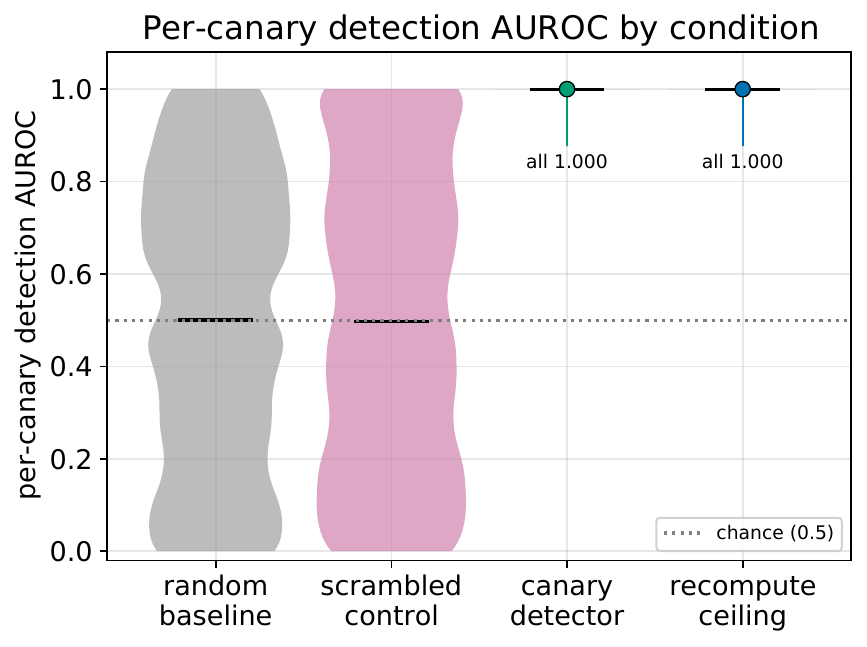}
  \caption{Detector versus controls: per-canary $A^{(t)}$
  from~\eqref{eq:auroc_canary} pooled over the factorial. The detector and the
  recompute ceiling concentrate at 1.0; the random baseline and the
  scrambled-label control spread around the 0.5 chance line.}
  \label{fig:detectors}
\end{subfigure}
\caption{Mechanism and detector performance. Higher is better in both panels.}
\label{fig:mech-det}
\end{figure}

\begin{figure}[t]
\centering
\begin{subfigure}[b]{0.60\linewidth}
  \centering
  \includegraphics[width=\linewidth]{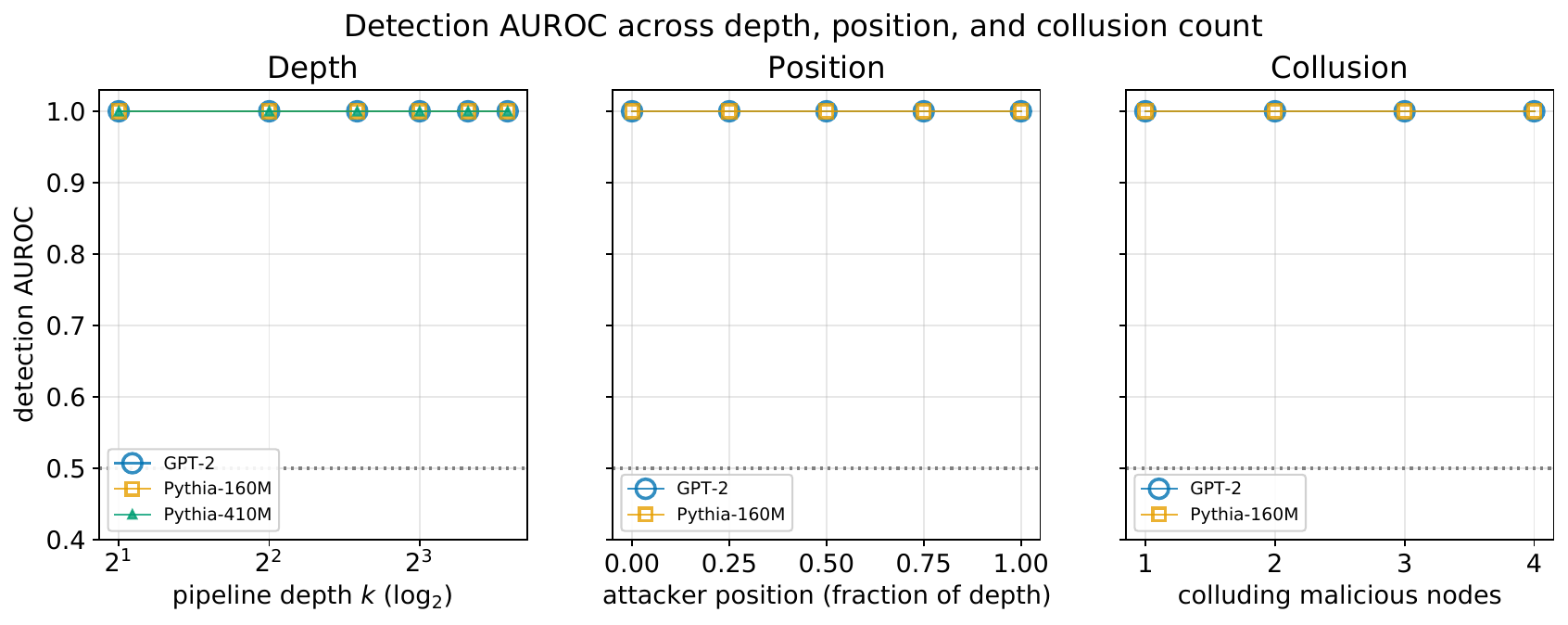}
  \caption{Invariance to adversary-controlled axes.}
  \label{fig:robustness}
\end{subfigure}\hfill
\begin{subfigure}[b]{0.375\linewidth}
  \centering
  \includegraphics[width=\linewidth]{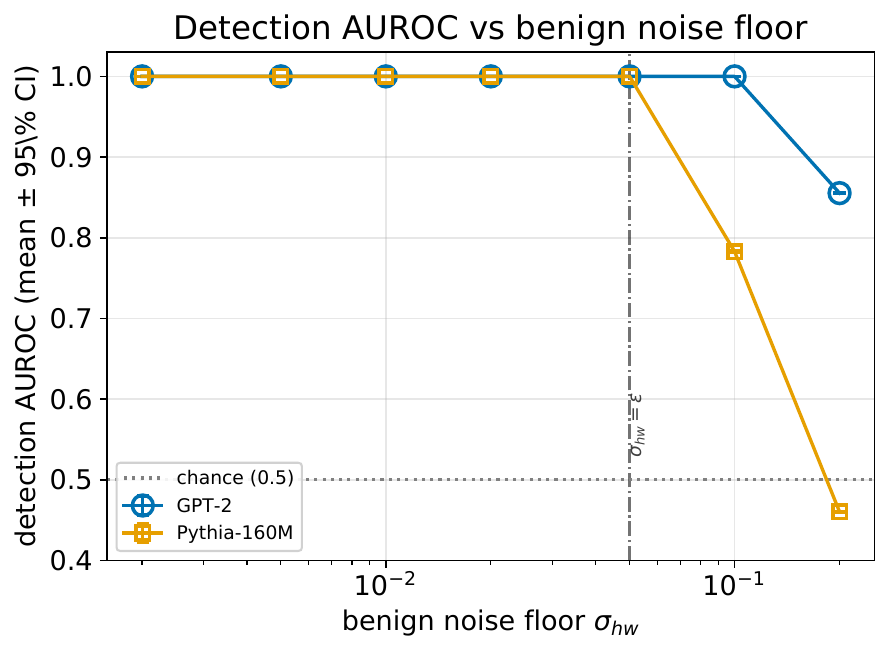}
  \caption{Detection breaking point.}
  \label{fig:snr}
\end{subfigure}
\caption{Robustness and breaking point; $\widehat{A}$ higher is better.
(a)~$\widehat{A} = 1.000$ is invariant to pipeline depth $k$ including a 24-block
model (left), attacker position (center), and $c \in \{1,2,3,4\}$ colluders
(right); dotted line 0.5 chance. (b)~$\widehat{A}$ against $\sigma_{hw}$ on a log
axis (mean and 95\% CI, three seeds) holds at 1.0 until $\sigma_{hw}$ approaches
$\epsilon = 0.05$, then falls toward chance, Pythia-160M first.}
\label{fig:robust-snr}
\end{figure}

This substudy raises the benign noise floor until the detector breaks, which
tests the separation condition $\sigma_{hw} \ll \epsilon$ (Figure~\ref{fig:mechanism}
shows the margin on which the condition depends; Figure~\ref{fig:snr} shows where it
closes). First, at $\sigma_{hw} = 0.1$, Pythia-160M falls to $\widehat{A} = 0.78$
while GPT-2 still holds at 1.000. Second, at $\sigma_{hw} = 0.2$, GPT-2 falls
to 0.86 and Pythia-160M reaches 0.46, near chance. Nothing collapses abruptly. As
$\sigma_{hw}/\epsilon \to 1$ the benign and malicious mismatch distributions
overlap, and $\mathbb{E}[A^{(t)}] \to 0.50$ by symmetry. Detection still works
when the benign noise is as large as the tamper, and it only starts to fail once the
noise is twice the tamper. There is more room here than $\sigma_{hw} \ll \epsilon$
suggests, and real runs sit well inside it: the noise level we model,
$\sigma_{hw} = 0.002$, is 25 times below the anchor tamper, and the realized
mismatch floor is lower still.

\subsection{Frontier substudy}
\label{subsec:res-frontier}

\begin{figure}[t]
\centering
\includegraphics[width=0.70\linewidth]{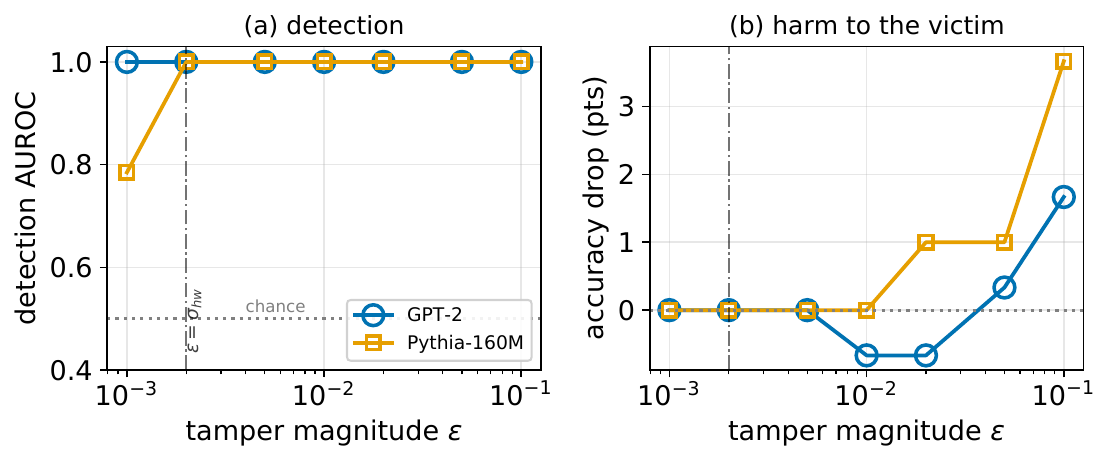}
\caption{Evasion frontier, at the fixed floor $\sigma_{hw} = 0.002$. (a)
detection $\widehat{A}$ and (b) the attacker's harm, the last-token accuracy
drop, both against $\epsilon$ on a log axis. The dash-dotted line marks
$\epsilon = \sigma_{hw}$. $\widehat{A}$ stays at 1.0 until $\epsilon$ falls
below $\sigma_{hw}$, where harm has already vanished; the small negative drops
in (b) are sampling noise around zero harm.}
\label{fig:frontier}
\end{figure}

The last substudy lowers $\epsilon$ toward the fixed floor $\sigma_{hw} = 0.002$
and tracks the detection and accuracy harm together (Figure~\ref{fig:frontier}).
First, detection holds: $\widehat{A} = 1.000$ all the way down to $\epsilon = 0.002$,
twice the floor. Second, it gives way at $\epsilon = 0.001$, below the floor,
dropping to 0.78 on Pythia-160M. By then the attack does no damage. At
$\epsilon = 0.001$ the answer loses 0.0 accuracy points. The attacker has to reach
$\epsilon = 0.1$, a hundred times larger, before the answer is somehow impacted. However, at this value, it
loses 1.7 to 3.7 points. Our method can catch every attack with such a magnitude. Therefore, the attacker has to
make a decision on either damage the answer, or stay hidden.

\begin{table}[t]
\begin{minipage}[t]{0.545\linewidth}
\centering
\ExplSyntaxOn\dim_set:Nn \l_tbl_width_dim {\linewidth}\ExplSyntaxOff
\caption{Main factorial (216 configurations, 100 canaries each). Detection AUROC
$\widehat{A}$ of the canary detector against the malicious node, with three
reference conditions. Higher is better; the baseline and negative control should
sit at 0.50 and the ceiling at 1.00.}
\label{tab:main}
\small
\setlength{\tabcolsep}{4pt}
\begin{tabular}{lcccc}
\toprule
Detector setting & $\widehat{A}$ & Base. & Neg.\ ctrl & Ceil. \\
\midrule
Full detector & 1.000 & 0.50 & 0.50 & 1.000 \\
Ablation: noisy reference & 1.000 & 0.50 & 0.51 & 1.000 \\
Ablation: cosine metric & 1.000$^{\dagger}$ & 0.50 & 0.50 & 0.865 \\
\bottomrule
\end{tabular}

\vspace{0.4em}
{\footnotesize $\dagger$ $\widehat{A} = 0.194$ on the 0.10 rescale; $\widehat{A}
= 1.000$ on the other five strategies.}
\end{minipage}\hfill
\begin{minipage}[t]{0.435\linewidth}
\centering
\ExplSyntaxOn\dim_set:Nn \l_tbl_width_dim {\linewidth}\ExplSyntaxOff
\caption{Harm versus detection by tampering strategy, full detector. Accuracy
drop is the fall in last-token accuracy in percentage points (higher is more
harmful). $\widehat{A} = 1.000$ regardless of the harm the strategy causes.}
\label{tab:harm}
\small
\setlength{\tabcolsep}{4pt}
\begin{tabular}{lcc}
\toprule
Tampering strategy & Acc.\ drop & $\widehat{A}$ \\
\midrule
Gaussian, $\epsilon = 0.02$ & 0.5 & 1.000 \\
Gaussian, $\epsilon = 0.05$ & 1.9 & 1.000 \\
Gaussian, $\epsilon = 0.10$ & 5.7 & 1.000 \\
Gaussian, $\epsilon = 0.20$ & 9.6 & 1.000 \\
Sign flip & 26.8 & 1.000 \\
Rescale, $\alpha = 0.10$ & 0.5 & 1.000 \\
\bottomrule
\end{tabular}
\end{minipage}
\end{table}

\section{Limitations and Future Work}
\label{sec:limitations}

\paragraph{Noise model fidelity}
The benign-node model in~\eqref{eq:benign} draws isotropic Gaussian noise at a
single level $\sigma_{hw}$, while real nondeterminism is structured: fp16
rounding is non-Gaussian, and operator reordering across GPU models produces
correlated discrepancies. Every result here rests on where the true floor sits
relative to a harmful $\epsilon$. Profiling
$\sigma_{hw}$ across a live heterogeneous pool, consumer GPUs on
Petals~\cite{petals2023} being the obvious target, would replace the modeled
floor with a measured one and re-run the signal-to-noise substudy against it.

\paragraph{Model scale}
All experiments use models of 124M to 410M parameters, while BLOOM-176B on
Petals~\cite{petals2023} is much larger, with wider hidden
states and different activation statistics. Detection may saturate more slowly
there, or the smallest detectable $\epsilon$ may rise above the smallest harmful
one, which would invert the frontier result of
Section~\ref{subsec:res-frontier}. Tracking $\widehat{A}$, the effective floor,
and the minimum harmful $\epsilon$ up through Pythia-2.8B and BLOOM-7B would
show whether the margin is a property of the mechanism or of small models.

\paragraph{Adaptive adversaries}
Our attacker is fixed in two ways, and each omission is its own optimization
problem. It commits to $\epsilon$ and $\tau$ before deployment and never
observes the detector, so an attacker that probes $\sigma_{hw}$ first and then
sets $\epsilon$ just beneath it is untested. It also tampers on every query, so
an attacker that fingerprints the canary distribution $\mathcal{C}$ and corrupts
only real traffic is untested as well. Building both, one maximizing harm
subject to staying under the floor and one maximizing the distinguishability of
$\mathcal{C}$ from live traffic, is what would show whether the mutual exclusion
of Section~\ref{subsec:res-frontier} survives an adversary that adapts.

\section{Conclusion}

We studied whether a node in a peer-to-peer distributed LLM inference pipeline
can be caught tampering with the activations it forwards, without paying for a
second forward pass. Our answer transplants the known-answer canary trap onto
intermediate activations: a client mixes canaries whose correct activations it
holds in advance into ordinary traffic, measures each shard's relative-L2
departure from that reference, and ranks shards rather than thresholding them.
Across a pre-registered evaluation of 408 configurations, the detector reaches
AUROC 1.000 on the 72 factorial configurations in which the full mechanism runs,
covering both architectures, both datasets, and all six attacks, a gain of 0.500
over the random baseline that clears the pre-registered 0.40 threshold. The
scrambled-label control collapses to chance, so the signal comes from the canary
mechanism and not from the measurement setup, and the redundant-recompute
ceiling also reaches 1.000, which is the expensive guarantee this detector
matches at one activation read.

The method works because the gap it measures is wide. The malicious shard's mismatch
tracks the tamper magnitude while an benign shard's tracks the float noise
floor, and at the smallest magnitude we tested those differ by 22 times, rising
past 2000 times under sign flip. A margin that size is beyond the reach of
benign drift, which is why detection needs no calibrated cutoff and why it
survives the axes an adversary actually controls: $\widehat{A}$ stays at 1.000
across pipeline depths from 2 to 12 blocks, at every attacker position, and with
up to four colluding nodes. Each shard is judged against its own stored
reference, not against a majority vote, so a growing malicious minority does not
erode the test.

The same margin bounds the result, and the substudies locate where it runs out.
Detection survives a noise floor as large as the tamper itself and only fails
beyond that: on Pythia-160M $\widehat{A}$ falls to 0.78 once the floor is twice
the tamper magnitude, and to 0.46 at four times. Shrinking the tamper below the
floor evades the detector too, but at $\epsilon = 0.001$ that attack costs the
victim zero accuracy points, which leaves an adversary with a choice rather than
an opening.

\bibliography{references}
\end{document}